\documentclass[twocolumn,showpacs,preprintnumbers,amsmath,amssymb,superscriptaddress]{revtex4-1}

\usepackage{graphicx}
\usepackage{dcolumn}
\usepackage{bm}
\usepackage{tabularx}
\usepackage{ulem} 
\newcolumntype{M}{>{\centering\arraybackslash}m{1.85cm}}
\usepackage[export]{adjustbox}
\usepackage{float}
\usepackage{color}   
\usepackage{hyperref}
\hypersetup{
    colorlinks=true, 
    linktoc=all,     
    linkcolor=blue,  
}

\makeatletter
\newcommand{\colorcaption}[2][]{%
  \begingroup%
  \renewcommand{\@caption@fignum@sep}{ (Color online). }%
  \caption[#1]{#2}%
  \endgroup%
}
\makeatother
\begin{document}

\title{Prediction of the neutron drip line in oxygen isotopes using quantum computation}

\author{ Chandan Sarma }
\email{c$_$sarma@ph.iitr.ac.in}
\address{Department of Physics, Indian Institute of Technology Roorkee, Roorkee 247667, India}	
\author{Olivia Di Matteo }
\email{olivia@ece.ubc.ca}
\address{Department of Electrical and Computer Engineering, The University of British Columbia, Vancouver, British Columbia V6T 1Z4, Canada}
\author{Abhishek Abhishek}
\email{abhiabhi@student.ubc.ca}
\address{Department of Electrical and Computer Engineering, The University of British Columbia, Vancouver, British Columbia V6T 1Z4, Canada}
\author{Praveen C. Srivastava\footnote}
\email{Corresponding author: praveen.srivastava@ph.iitr.ac.in}
\address{Department of Physics, Indian Institute of Technology Roorkee, Roorkee 247667, India}

\date{\hfill \today}


\begin{abstract}
In the noisy intermediate-scale quantum era, variational algorithms have become a standard approach to solving quantum many-body problems. Here, we present  variational quantum eigensolver (VQE) results of selected oxygen isotopes within the shell model description. The aim of the present work is to locate the neutron drip line of the oxygen chain using unitary coupled cluster (UCC) type ansatze with different microscopic interactions (DJ16, JISP16, and N3LO), in addition to a phenomenological USDB interaction. While initially infeasible to execute on contemporary quantum hardware, the size of the problem is reduced significantly using qubit tapering techniques in conjunction with custom circuit design and optimization. The optimal values of ansatz parameters from classical simulation are taken for the DJ16 interaction, and the tapered circuits are run on IonQ's Aria, a trapped-ion quantum computer. After applying gate error mitigation for three isotopes, we reproduced exact ground state energies within a few percent error. The post-processed results from hardware also clearly show $^{24}$O as the drip line nucleus of the oxygen chain. Future improvements in quantum hardware could make it possible to locate drip lines of heavier nuclei.

\end{abstract}


\maketitle
\newpage
\section{Introduction}
Atomic nuclei are complex many-body systems composed of nucleons interacting via strong nuclear force. Understanding nuclear properties from the nucleon-nucleon force is one of the main goals of low-energy nuclear physics. Like other quantum many-body problems, the structure of atomic nuclei can be effectively solved using configuration-interaction methods. One such method that is very successful for solving many-body problems of nuclear structure is the nuclear shell model \cite{SM1, SM2, SM3}. But, the exponential increase in Hilbert space with increasing nucleon numbers has become a computational challenge for classical computers. Quantum computers are emerging as promising tools for solving many-body problems across the spectrum of physical sciences. These devices are natural quantum systems in which the principles of quantum mechanics, like the superposition principle and entanglement, are embedded. 

In the present noisy intermediate-scale quantum (NISQ) era \cite{nisq}, variational methods like the variational quantum eigensolver (VQE) \cite{vqe} are among the most successful quantum algorithms. The VQE is a hybrid classical-quantum algorithm exploiting the benefits of quantum computing for state preparation and measurements, and the benefits of classical computers for optimization. While this approach is widely used in quantum chemistry, there are comparatively fewer applications in nuclear physics \cite{QC_n1, QC_n2, graycode, pooja1, pooja2, pooja3, QC_n3, QC_n4, QC_n5, QC_n6, QC_n7, QC_n8, QC_n9}.  Our present work is based on \cite{QC_n3}, where some of the nuclei from $p$- and $sd$-shell are studied, including $A$ = 20, 22 isotopes of oxygen using unitary coupled cluster (UCC) ansatz. We are extending the work to include the whole chain of even-even oxygen isotopes from $N$ = 10 to 18 to evaluate the ground state energies using UCC ansatz. By doing so, our intention is to establish $^{24}$O as the neutron drip line nucleus of the oxygen chain. Due to its semi-magic nature, the oxygen isotopic chain has been the testing ground of different theoretical nuclear approaches for a long time. More than twenty years ago, $^{24}$O was established to be the heaviest bound nucleus of the oxygen chain \cite{dripline1}. This isotope at the neutron drip line can be understood theoretically as single nucleons filling mean-field single-particle orbitals \cite{dripline2}. Recently, the drip line of fluorine and neon isotopic chains was confirmed in RIKEN Radioactive Isotope Beam Factory \cite{dripline3}. Tsunoda \textit{et al}. in \cite{dripline2} show that the neutron drip line from fluorine ($Z$ = 9) to magnesium ($Z$ = 12) can be predicted in terms of deformation mechanism. The discovery of $^{39}$Na, the most neutron-rich sodium nucleus observed so far is reported in Ref. \cite{39Na}.
As the drip line prediction for larger systems becomes computationally and experimentally intractable, novel approaches, such as quantum computing, must be pursued.

In the present work,  quantum computing is used to locate the drip line nucleus for the oxygen chain. VQE calculations are applied to a variety of different phenomenological and microscopic interactions. While straightforward to verify through simulation, we also run the quantum circuits of different isotopes on quantum hardware at the variational minimum. The hardware results for $^6$Li \cite{QC_n5} show that the error ratio is an important factor to consider while running variational algorithms on quantum hardware. Using qubit tapering methods, optimized circuits, and a custom transpilation tool to significantly reduce running costs and errors, we obtain ground state energies of all five even-A oxygen isotopes. Successful implementation of quantum hardware results for the prediction of the oxygen drip line opens the path for other nuclear isotopic chains. In the long term, the improvement of quantum hardware and the development of sophisticated nucleon-nucleon interactions would revolutionize this area of nuclear structure.

\section{Theoretical Framework}
 The nuclear shell model is a very successful theory for nuclear structure. This approach utilizes large-scale diagonalization in a many-body harmonic oscillator basis to explain several low-energy structural properties of nuclei. As the nuclear force is rotationally invariant, single particle harmonic oscillator states having well-defined quantum numbers, $n$, $l$, $j$, and $j_z$, are very good choices for constructing many-body states. Furthermore, nuclear force is the same for both protons and neutrons to a very good approximation resulting in additional quantum numbers, isospin ($t = 1/2$), and the third component of isospin ($t_z = \pm 1/2$).  Though the number of single-particle states could typically be small, the total number of many-particle states increases rapidly with the increase in the number of nucleons. The total angular momentum $J$ and total isospin $T$ are good quantum numbers of many-body nuclear states. The third components of ($J$, $T$),  ($M$, $T_z$), represent the addition of $j_z$ and $t_z$ of each nucleon in a nucleus, and are also good quantum numbers. The $M$-scheme is one of the preferred ways of constructing many-body states for nuclear shell model having a well-defined $M$, and some of the prominent shell model codes like Antoine \cite{antoine}, NuShellX \cite{nushellx}, KShell \cite{kshell}, and Bigstick \cite{Bigstick} use this scheme.

\subsection{Hamiltonian}
In this work, we considered a shell model description of even-A oxygen isotopes having (A-16) valence neutrons added to the inert $^{16}$O core. The neutrons for different O-isotopes lie in the $sd$-model space comprising 0$d_{5/2}$, 1$s_{1/2}$, and 0$d_{3/2}$ harmonic oscillator orbitals. Apart from the phenomenological USDB interaction \cite{usdb}, we also use some of the recently developed effective microscopic interactions for this work: JISP16 \cite{sd_int1}, N3LO \cite{sd_int1}, and DJ16 \cite{sd_int2}. These effective interactions for $sd$-space are derived from the original interactions \cite{jisp16, n3lo, dj16} using the $ab$ $initio$ no-core shell model (NCSM) \cite{ncsm_2013,priyanka1,priyanka2,chandan} wave function and Okubu-Lee-Suzuki (OLS) \cite{ols_1994} technique. 
These microscopic interactions are recently applied for upper $sd$ shell nuclei in Refs. \cite{priyanka_EPJA,priyanka_NPA,subh_NPA}.
The single particle energies (SPE) used for different interactions are mentioned in \autoref{T1}.

\begin{table}[H]
	\centering
	\caption{The single-particle energies of $sd$ orbitals for different interactions are shown in MeV.}
	\begin{tabular}{|c|c|c|c|c|}
		\hline
		Single Particle States & \hspace{1.75mm} USDB  & \hspace{1.75mm} DJ16   & \hspace{1.75mm} JISP16  & \hspace{1.75mm} N$^3$LO  \\
		\hline
		0d$_{5/2}$ & -3.9257 & -3.302 &  -2.270 & -3.042\\
		1s$_{1/2}$ & -3.2079 & -3.576 &  -3.068 & -3.638\\
		0d$_{3/2}$ & 2.1117 &  6.675 &  6.262 & 3.763\\
		\hline
	\end{tabular}
	\label{T1}
\end{table}

The shell model Hamiltonian in the second quantization is written as
\begin{equation}
\label{eq1}
H = \sum_{i}\epsilon_i \hat{a}^\dagger_i  \hat{a}_i + \frac{1}{2} \sum_{i, j, k, l} V_{ijlk} \hat{a}^\dagger_i \hat{a}^\dagger_j \hat{a}_k \hat{a}_l.
\end{equation}
Here, $\hat{a}^\dagger_i$ and $\hat{a}_i$ are the creation and annihilation operators of a nucleon in state $|i\rangle$. The coefficients $\epsilon_i$ and $V_{ijlk}$ are the single-particle energies and two-body matrix elements (TBMEs). The nucleon state can be represented by $|i\rangle$ = $|n, l, j, j_z, t_z\rangle$, where $n$ and $l$ are the radial and angular momentum quantum numbers, respectively. Total spin $j$ = 5/2, 3/2, and 1/2 for $sd$-space and $j_z$ and $t_z$ are projections of spins and isospins. For the $sd$-model space, we need 12 states for protons and 12 states for neutrons. As we are working on oxygen isotopes having a magic number of protons, we will need only 12 states for valence neutrons. We can represent this problem in the quantum computation framework using $N$ = 12 qubits. The qubit representation of neutron states is given in \autoref{T2}.

\begin{table}[H]
	\caption{Qubit representation of the orbitals of $sd$-model space. As we are dealing with the neutrons alone, $t_z$ = 1/2 for each qubit.}
	\centering
	\begin{tabular}{|c|c|c|c|c|c|c|c|c|c|c|c|c|}
		\hline
		Qubit & \hspace{1.0mm} 0 \hspace{1.0mm} & \hspace{1.0mm} 1 \hspace{1.0mm}&\hspace{1.0mm} 2 \hspace{1.0mm}&\hspace{1.0mm} 3 \hspace{1.0mm}& \hspace{1.0mm}4 \hspace{1.0mm}& \hspace{1.0mm}5 \hspace{1.0mm}& \hspace{1.0mm}6 \hspace{1.0mm}&\hspace{1.0mm} 7 \hspace{1.0mm}&\hspace{1.0mm} 8 \hspace{1.0mm}&\hspace{1.0mm} 9 \hspace{1.0mm} & \hspace{.7mm} 10 \hspace{.7mm} & \hspace{.7mm} 11 \hspace{.7mm} \\
         \hline
         $n$ & 0 & 0 & 0 & 0 & 0 & 0 & 1 & 1 & 0 & 0 & 0 & 0\\
         \hline
         $l$ & 2 & 2 & 2 & 2 & 2 & 2 & 0 & 0 & 2 & 2 & 2 & 2 \\
         \hline
		2$j$ & 5 & 5 & 5 & 5 & 5 & 5 & 1 & 1 & 3 & 3 & 3 & 3 \\
		\hline
		2$j_z$ & -5 & 5 & -3 & 3 & -1 & 1 & -1 & 1 & -3 & 3 & -1 & 1 \\
		\hline
	\end{tabular}
 \label{T2}
\end{table}

 The shell model Hamiltonian is converted into the qubit Hamiltonian with the Jordan-Wigner (JW) transformation using the mapping
\begin{eqnarray}
\hat{a}^\dagger_k = \frac{1}{2}  \left ( \prod_{j = 0}^{k-1} -Z_j \right)(X_k -iY_k), \label{eq2}\\
\hat{a}_k = \frac{1}{2}  \left ( \prod_{j = 0}^{k-1} -Z_j \right)(X_k + iY_k). \label{eq3} 
\end{eqnarray}
\noindent Here, $X_k$, $Y_k$, and $Z_k$ are the Pauli matrices applied to $k^{th}$ qubit. Using these relations, the shell model Hamiltonian is re-expressed in terms of Pauli strings. The empty and occupied single-particle states are represented by $|0\rangle$ and $|1\rangle$, respectively.

All four interactions we are working with are usually available in the $J$-scheme with 63 two-body matrix elements, which are difficult to use in quantum computation directly as angular momentum couplings between different nucleon states make the problem more complicated.
To make the problem more straightforward, a transformation to the $M$-scheme is applied, yielding a Hamiltonian in the second-quantized form as in \autoref{eq1}. The relevant equations for the transformation are included in Appendix A, as well as in our code repository \cite{github}. The transformed TBMEs in $M$-scheme can be divided into three parts: proton-proton ($pp$), neutron-neutron ($nn$), and proton-neutron ($pn$). As we are working on the oxygen chain having valence neutrons only, we transform the $J$-scheme matrix elements to the $nn$ $M$-scheme TBMEs.  Using the $M$ scheme TBMEs and the creation and annihilation operators defined in \autoref{eq2} and \autoref{eq3}, a second-quantized Hamiltonian is constructed in terms of Pauli operators. This JW-transformed Hamiltonian contains 1611 Pauli terms, out of which only a limited number are useful for measurements; in the next sections, we discuss how the Hamiltonian is further processed to reduce its size.

\subsection{Initial State Preparation}

\begin{figure*}
    \centering
    \includegraphics[width=0.8\textwidth]{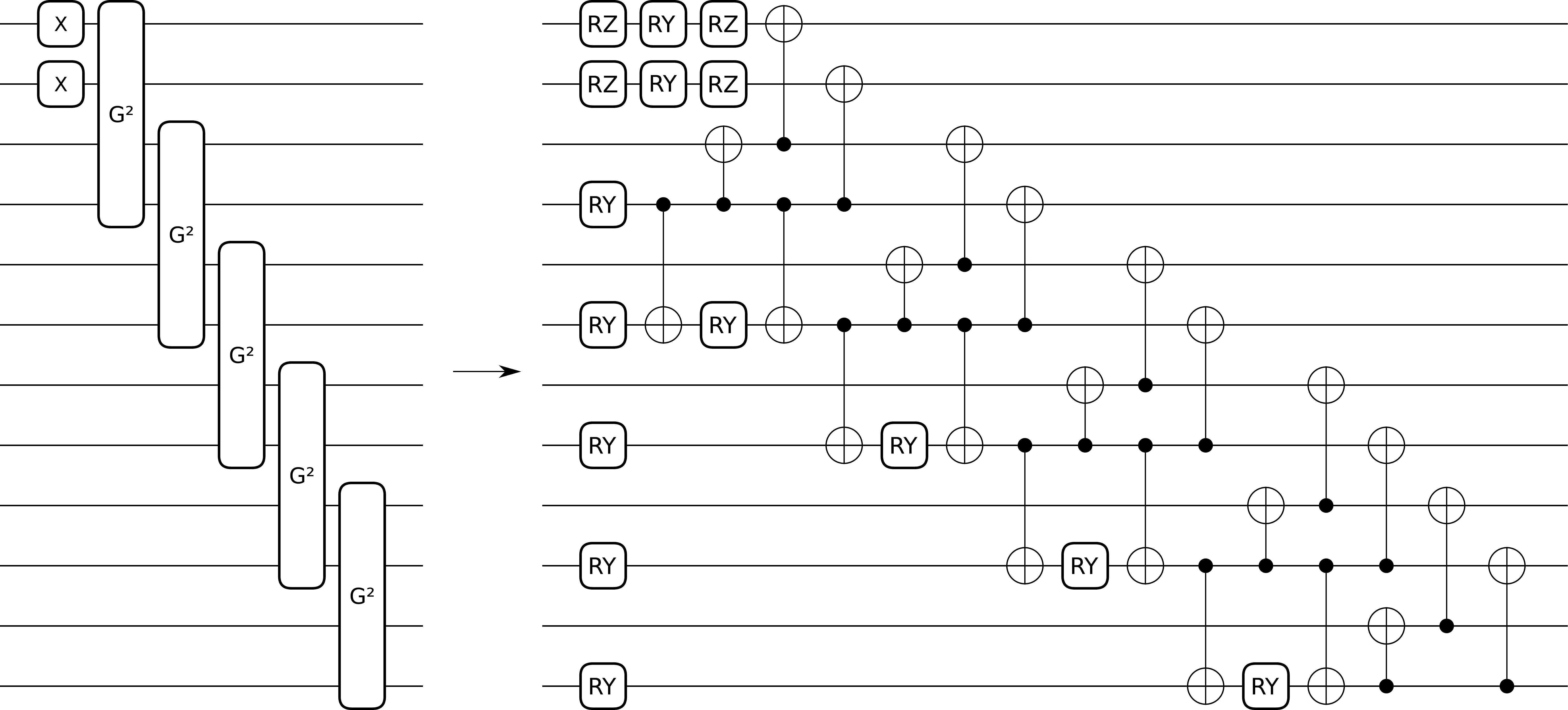}
    \caption{Ground state construction for $^{18}$O. (Left) Structure of double excitations required to prepare the ground state. (Right) Custom, highly-optimized circuit to prepare the same ground state that leverages the fact that we start in the all-zeros state, composed primarily of $RY$ and $CNOT$ gates. Further optimization, e.g., removal of initial $RZ$ gates, may be performed as needed with subsequent transpilation and optimization passes.} 
    \label{fig:o18-12qubit-version}
\end{figure*}

The initial state preparation is the minimum starting point in finding out the ground state energies using VQE approaches. Unlike other work, which uses the Hartree-Fock state as the reference state, here we choose the lowest-energy Slater determinant. In JW mapping, the occupied orbitals of the reference state are represented by applying a suitable number of $X$ gates. Consider, for example, the case of $^{18}$O ground state having spin $J$ = 0. The lowest-energy Slater determinant is $|0, 1 \rangle$ for all our considered interactions with $M$ = 0. This state can be represented in a computational basis as:
\begin{equation}
	|110000000000 \rangle = X_0X_1|000000000000 \rangle. \label{eq:initial_state}
\end{equation}

Similarly, we can represent the reference states corresponding to the ground state of other oxygen isotopes. Apart from the ground state, energies of some of the low-lying excited states can also be calculated using the VQE method. The initial state preparation of the excited states is just like that for the ground states. For example, the first excited state of $^{18}$O has spin $J$ = 2 which corresponds to $M$ = -2, -1, 0, 1, or 2. We can ideally select any combination of two qubits that result in these five values of $M$. Out of these possible combinations, we select $|1, 4 \rangle$ as the reference state having $M$ = 2 which lies in a subspace orthogonal to the ground state. The details of the circuit design and ideal simulator results for the first excited state of $^{18}$O is given in Appendix B.

\subsection{Variational Ansatz}
\label{VA}

The VQE is a quantum algorithm that can be used to determine the ground state of a qubit Hamiltonian. It is well-described in many other works (see, e.g., \cite{vqe, vqa_review, quantum_chem}), so it is outlined here only at a high level.

In the VQE, a register of qubits is prepared in an initial state, e.g., as described in the previous section. A quantum computer then applies a sequence of operations based on real-valued parameters (a parametrized circuit, or circuit ansatz), and the expectation value of the problem Hamiltonian is measured. The measurement results are input to an optimization algorithm running on a classical computer. The optimizer uses the expectation value as its objective function, and seeks the set of gate parameters that minimizes it. Assuming a suitable ansatz is used, the ground state will be prepared by the quantum computer by running the circuit at the optimal parameter values.

For the successful implementation of quantum computation methods like the VQE, the construction of an ansatz is one of the crucial steps. The unitary coupled cluster ansatz (UCC) is the most commonly used ansatz for finding ground states in quantum chemistry and nuclear physics. The UCC involves cluster operators that act on a reference state $|\psi_0\rangle$,
\begin{equation}
|\psi(\theta)\rangle = e^{i(\hat{T} (\theta) - \hat{T}^\dagger (\theta))} |\psi_0\rangle.
\end{equation}

These operators can be decomposed into singles, doubles, etc., and excitation operators that drive occupied orbitals to unoccupied ones,
\begin{align}
\hat{T} = \hat{T_1} + \hat{T_2} + ..., \\
 \hat{T_1} = \sum_{i, \alpha}\theta^\alpha_i a^\dagger_i a_\alpha, \\ \hat{T_2} = \sum_{ij, \alpha\beta}\theta^{\alpha \beta}_{ij} a^\dagger_i a^\dagger_j a_\alpha a_\beta,
\end{align}

\noindent where $\alpha, \beta$ are occupied states and $i, j$ are unoccupied states. 

In this work, we are using a simple unitary coupled-cluster doubles (UCCD) type ansatz for the ground state of even-A oxygen isotopes given in \cite{universal_gates}. The quantum circuits are designed using double and controlled-double excitations expressed in terms of Givens rotation $G^{(2)}$ acting on a reference state \cite{universal_gates}. Only pair-wise excitations within the same orbital, or from one orbital to the other, are applied, which conserves $j_z$.
The $M$-scheme dimensions of the ground states of these oxygen isotopes and the number of Slater determinants (SDs) in each ansatz are given in the \autoref{T3}, along with the number of parameters needed for each ansatz inside brackets.

\begin{table}[H]
	\caption{The $M$-scheme dimensions of the ground states of different oxygen isotopes are mentioned along with the no. of Slater determinants (SD) considered in the construction of ansatze. The required number of parameters is shown in the bracket.}
	\centering
	\begin{tabular}{| c | c | c |}
		\hline
		Isotopes & $M$-scheme dimension & No. of SDs in ansatz\\
		\hline 
	    $^{18}$O	& 14 & 6 (5) \\
		$^{20}$O	& 81 & 15 (14) \\
		$^{22}$O	& 142 & 20 (19)\\
		$^{24}$O	&  81 & 15 (14)\\
	    $^{26}$O	& 14  & 6 (5)\\
		\hline
	\end{tabular}
	\label{T3}
\end{table}

A circuit that constructs the ground state of $^{18}$O using double excitations is shown in the left panel of \autoref{fig:o18-12qubit-version}. Each double excitation comprises a single variational parameter. In order to run this circuit on hardware, the double excitations must be decomposed into 1- and 2-qubit gates. Naively applying the decomposition circuit for a double excitation results in a very inefficient construction; instead, we design a custom circuit that leverages the fact that we start with the single basis state in \autoref{eq:initial_state}, and uses controlled $RY$ gates and $CNOT$s to achieve a ground state with the desired structure (a linear combination of 6 computational basis states, each with two sequential qubits in the $\vert 11 \rangle$ state). The same circuit can be used to construct the ground state of $^{26}$O, with the addition of a Pauli $X$ gate on each qubit at the end to exchange the role of 0s and 1s. Resource counts are shown in \autoref{tab:original}.

\begin{figure*}
    \centering
    \includegraphics[width=0.5\textwidth]{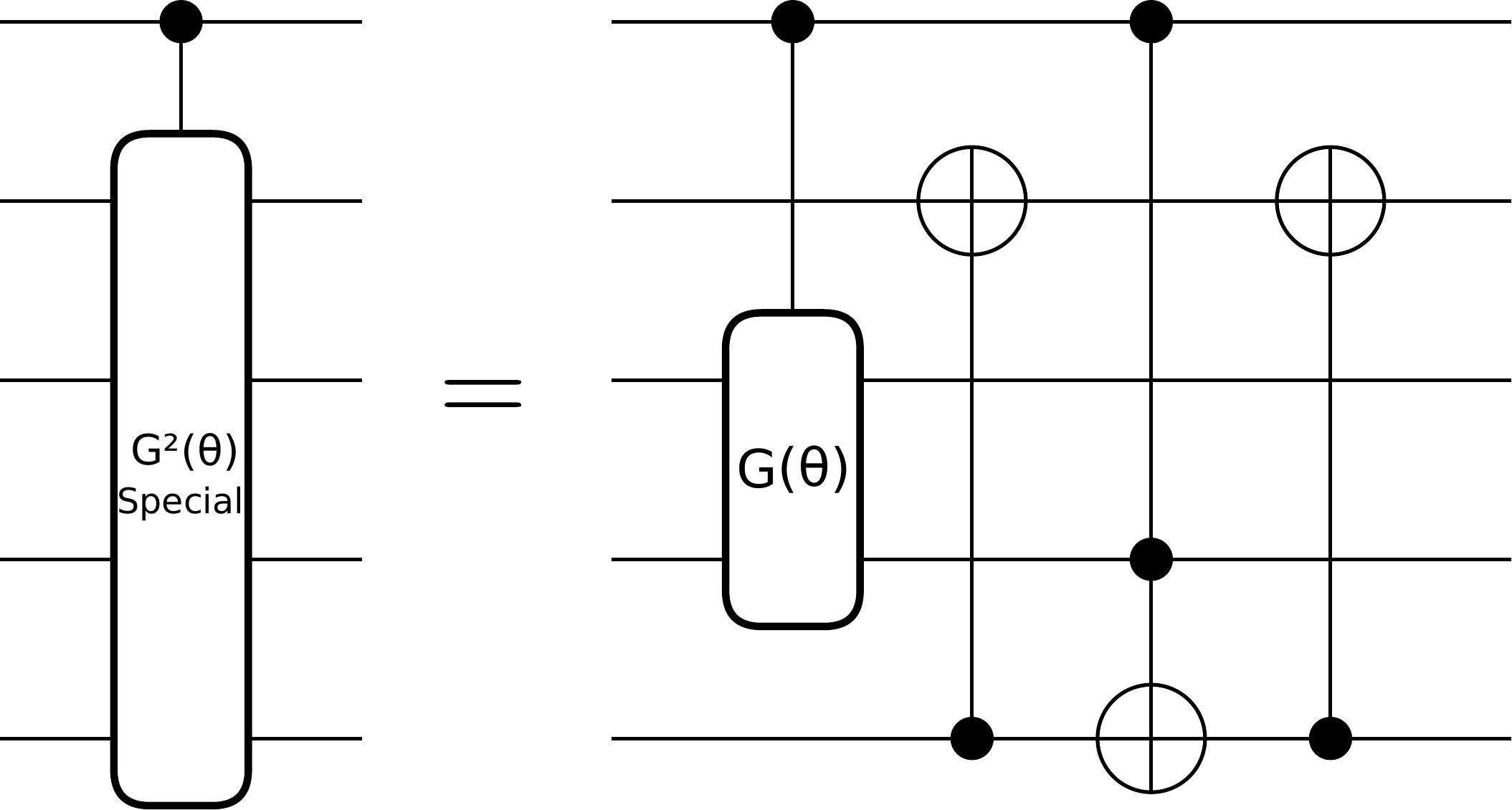}
    \caption{Decomposition of a special-purpose controlled double excitation. It can be used in the construction of ansatze for $^{20}$O, $^{22}$O, and $^{24}$O as not all computational basis states are present. It performs a correct controlled double excitation when the target register is in state $|1100\rangle$, but will adversely affect other basis states. The controlled single excitation is a fully general one, and its decomposition is shown in \autoref{fig:ctrl_single_exc}.}
    \label{fig:ctrl_dble_decomp}
\end{figure*}

\begin{figure*}
    \centering
    \includegraphics[width=0.7\textwidth]{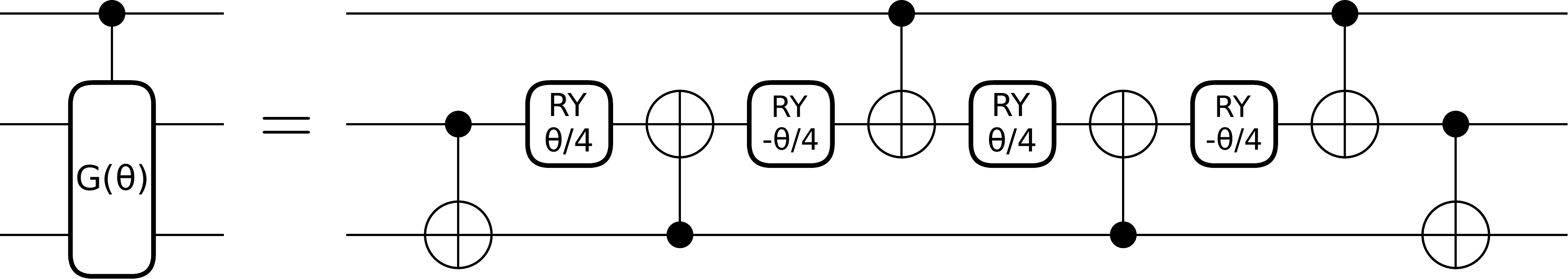}
    \caption{Decomposition of a controlled single excitation into $CNOT$ and $RY$ gates. This was discovered by applying gate identities similar to those described in \cite{elementary_gates}, followed by further reduction of $CNOT$ count using templates from \cite{templates}.}
    \label{fig:ctrl_single_exc}
\end{figure*}

For $^{20}$O, $^{22}$O, and $^{24}$O, the number of and nature of basis states involved in the ground state is such that singly- or even doubly-controlled double excitations are required. $^{20}$O and $^{24}$O have some symmetry (like $^{18}$O and $^{26}$O, we can use the same ansatz with a terminal layer of Pauli $X$), and each require 10 controlled-double excitations in addition to 4 double excitations. $^{22}$O requires 5 controlled doubles and 11 doubly-controlled doubles. Implementing these circuits requires a substantial amount of resources when applying naive decomposition strategies, so similar tricks as the $^{18}$O case were applied. In particular, due to the limited number of basis states involved in the initial parts of a circuit, a ``reduced" controlled double excitation with a simpler form can be used in many cases, instead of a full version. It is shown in \autoref{fig:ctrl_dble_decomp}, and requires a controlled single excitation, for which we found a simpler decomposition, shown in \autoref{fig:ctrl_single_exc}.

The first column in \autoref{tab:original} shows the resource counts for all our ansatze, following an additional optimization pass through the Qiskit transpiler with optimization level 3 \cite{qiskit}. Some of the 12-qubit circuits possess thousands of gates, far beyond the current limit of present-day quantum hardware, despite the design shortcuts and optimization. In what follows, we discuss the approaches used to simplify the problem to make hardware execution manageable.

\subsection{Qubit Tapering}
\label{subsec:tapering}

Out of the 1611 Pauli terms of the JW transformed $nn$ Hamiltonian, only 199 terms will be necessary due to the simple circuit design strategy discussed in \autoref{VA}.
So, instead of using the full Hamiltonian, we can use a reduced Hamiltonian containing 199 Pauli terms only.
All distinct Pauli observables can be grouped into disjoint sets of commuting terms, which can be measured simultaneously. There is more than one way of grouping them, such as by qubit-wise commutation having 13 sets or by commutation of the full observable, leading to 6 sets. For the 12-qubit quantum circuits discussed in \autoref{VA}, we use qubit-wise commutation measurements as the basis rotations are trivial to execute.

The size of the problem can be reduced using the symmetries of the JW transformed Hamiltonian \cite{qtap1, qtap2}. Using this method, which is implemented as part of PennyLane's quantum chemistry functionality \cite{pennylane, pennylane-qchem}, we can taper down the initial problem over 12 qubits to an equivalent  problem over 5 qubits. The associated Hamiltonian has only 52 terms instead of the original 199 terms. The Pauli terms of the tapered Hamiltonian can be grouped into 8 qubit-wise commuting sets or 6 general commuting sets. Though the qubit tapering reduces the size of the problem significantly, the number of variational parameters in each quantum circuit remains the same as the 12-qubit counterpart. 

\begin{figure*}
    \centering
    \includegraphics[width=0.4\textwidth]{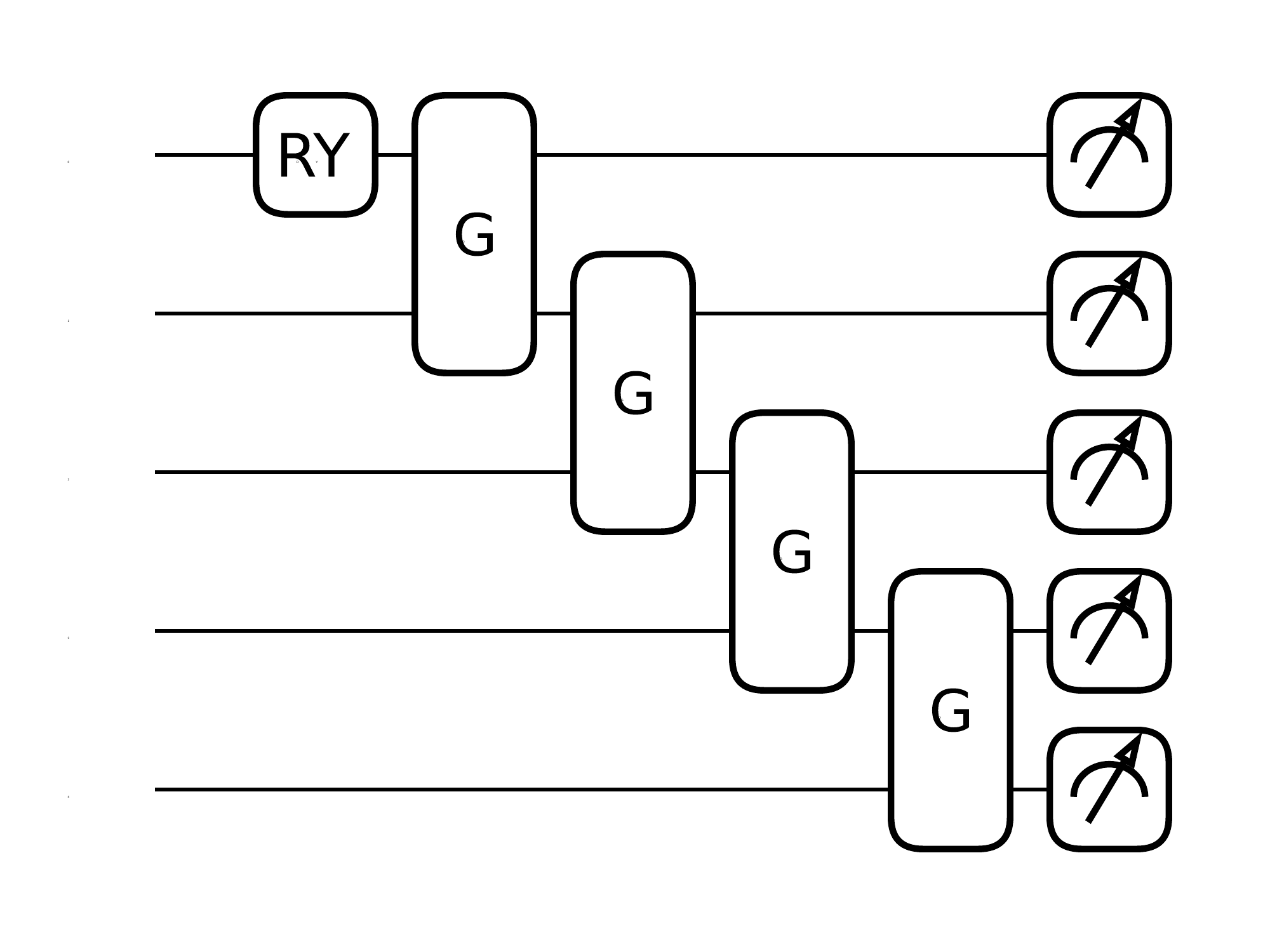}
    \includegraphics[width=\textwidth]{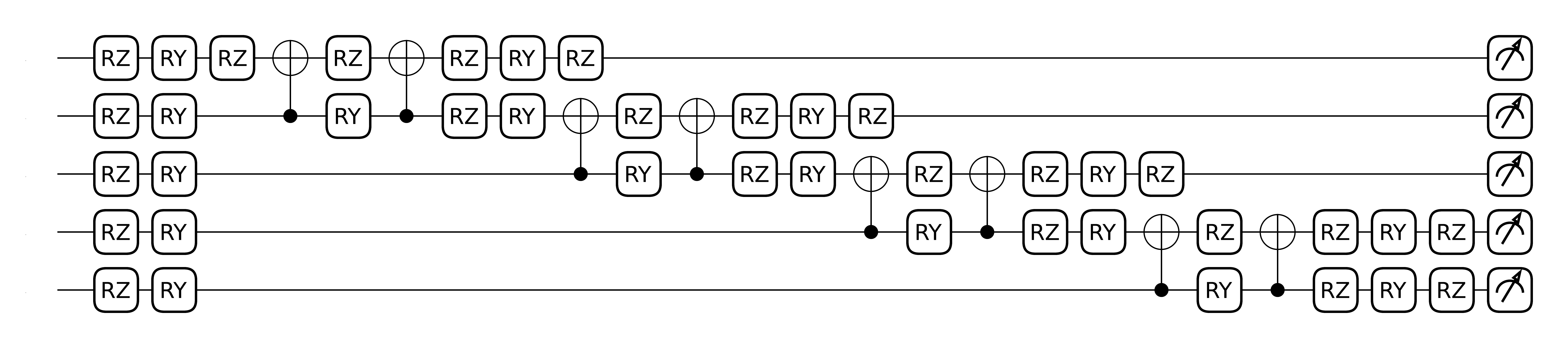}
    \caption{Circuits for $^{18}$O after tapering. Numerical values of parameters are omitted for clarity. (Upper) Overarching structure of the circuit in terms of parametrized single excitations, labelled by $G$. (Lower) The upper circuit decomposed into $RZ$, $RY$, and CNOT gates. This circuit is much more tractable than that of the original 12-qubit problem in \autoref{fig:o18-12qubit-version}.}
    \label{fig:18O_tapered_ansatz}
\end{figure*}

PennyLane's tapering implementation also produces the set of excitation gates that must be applied to the new system. However, similar to the 12-qubit case, we were able to further simplify these by leveraging: the fact that the system starts in the all 0s state and contains a limited number of basis states; a decomposition we found for single excitations that uses only 2 $CNOT$s; manual parallellization of some $CNOT$s; and a final pass through the Qiskit transpiler with optimization level 3. An example is shown in \autoref{fig:18O_tapered_ansatz}. The resource counts of the tapered circuits are presented in the final column of \autoref{tab:original}. One sees that with tapering, the problem becomes tractable on a near-term device.

\begin{table}
 \caption{Resource counts for original ansatze (12 qubits) and tapered ansatze (5 qubits).  All these circuits have been decomposed 
        and run through the Qiskit transpiler with optimization level 3, and are expressed in terms of $RY$, $RZ$ and $CNOT$ gates. 1Q and 2Q represent 1- and 2-qubit gate counts, and $d$ is the circuit depth.}
    \centering
    \label{tab:original}
   \begin{adjustbox}{width=0.38\textwidth}
    \begin{tabular}{|c||c|c|c||c|c|c|}
      \hline
      & & Original & & & Tapered & \\ \hline
      Iso. & 1Q & 2Q & $d$ & 1Q & 2Q & $d$\\ \hline
      18 & 13 & 23 & 15 & 40 & 8 &  24\\
      20 & 154 & 158 & 182 & 55 & 26 & 45 \\
      22 & 1063 & 787 & 1036 & 59 & 35 & 55 \\
      24 & 176 & 158 & 184 & 67 & 36 & 58 \\
      26 & 37 & 23 &  17 &  39 & 8 & 24 \\ \hline
    \end{tabular}
     \end{adjustbox}
\end{table}

\section{Results and Discussions}
The parametrized quantum circuits mentioned in \autoref{subsec:tapering} have been implemented on quantum simulators and trapped-ion quantum hardware. The classical simulation was implemented in PennyLane \cite{pennylane}, augmented by Qiskit's transpilation tools \cite{qiskit}; for the large 12-qubit circuits, computations were sped up with the JAX backend and just-in-time compilation \cite{jax}. The classical optimization process of the VQE method is performed using AdamOptimizer, which is a gradient-descent optimizer with an adaptive learning rate. The ground state energies of oxygen isotopes for different interactions are summarised in \autoref{tab:summary}. For each interaction, the simulated results for original and tapered quantum circuits are compared with the shell model diagonalization results. Our code and data is available on GitHub \cite{github}.\\ 

\subsection{Results from simulator}

\begin{figure}
   \centering
    \includegraphics[scale=0.61]{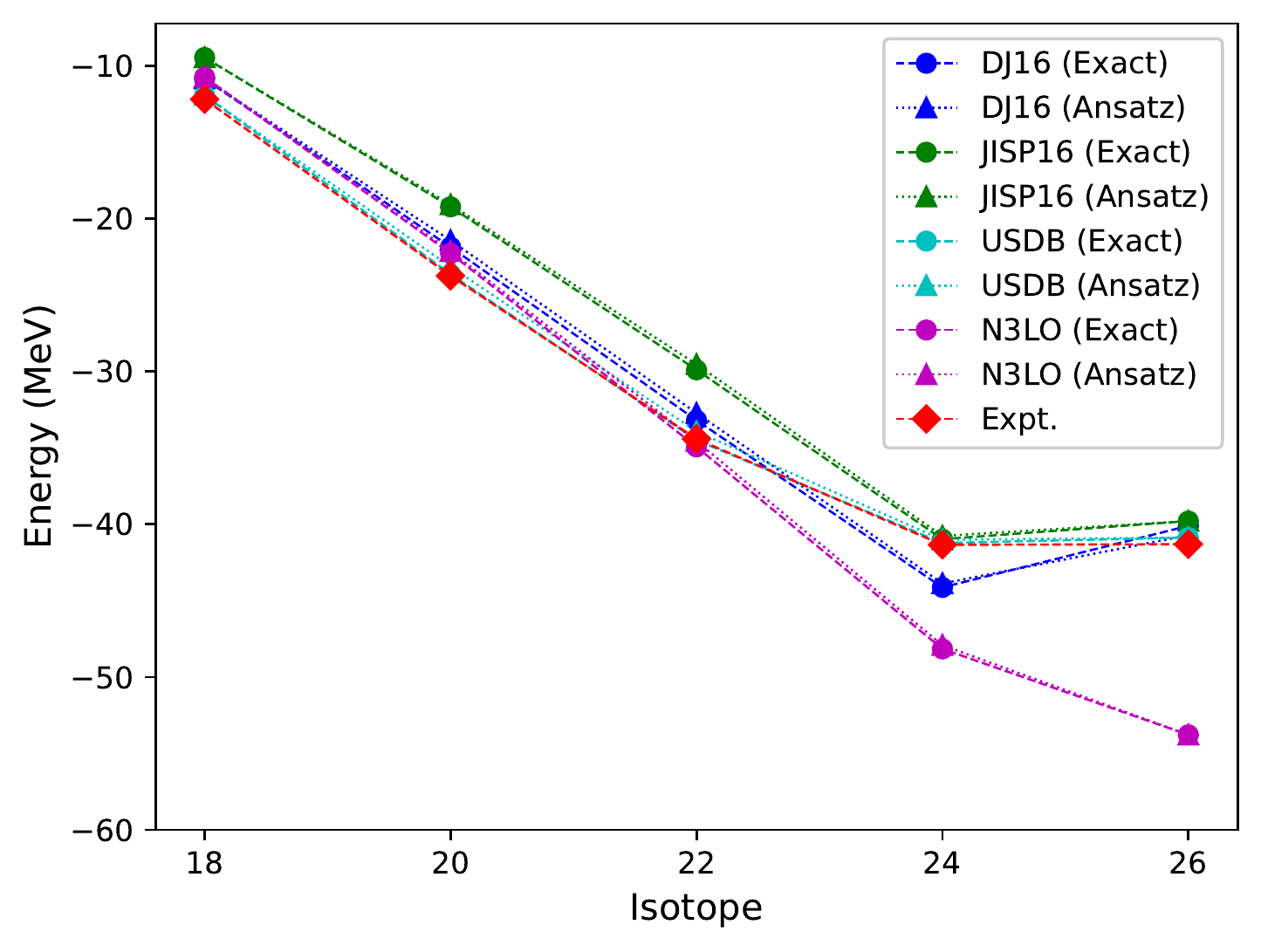}
    \caption{Theoretical ground state energies of oxygen isotopes relative to the $^{16}$O ground state are compared to the experimental data \cite{NNDC}. The shell model results for a particular interaction are shown as ``exact'' whereas the simulator results are shown as ``ansatz''. For the USDB interaction, mass dependence (18/(16+n))$^{0.3}$ is considered in the two-body matrix elements, while it is not considered for the microscopic interactions. The single particle energies for different interactions are mentioned in Table \ref{T1}.}
    \label{simulator_summary}
\end{figure}

The simplest system among the five oxygen isotopes we considered here is $^{18}$O having two neutrons in the $sd$ space. The ground state of this oxygen isotope is represented to a good approximation by a five-parameter quantum circuit representing six two-particle SDs. Without using fourteen SDs as represented by the $M$-scheme dimension in \autoref{T3}, we are able to reach energy with an error ratio less than 0.03 \% 
for the four interactions, we considered in this work. This result is comparable to the ground state energy obtained in Ref. \cite{QC_n6} using a five-layer ADAPT-VQE ansatz. The case of $^{26}$O is comparable, due to similarities in the circuit structure described in previous sections.

\begin{table*}
    \centering
        \caption{The ground state energies of even-$A$ oxygen isotopes are shown for all four interactions. For each interaction, the shell model diagonalization results are compared to the simulator results for the original 12 qubit circuits and the tapered ones with 5 qubits.}
    \begin{adjustbox}{width= \textwidth}
    \begin{tabular}{|c||c|c|c||c|c|c||c|c|c||c|c|c|}
      \hline
      & & USDB & & & DJ16 & & & JISP16 & & & N3LO & \\ \hline
      Iso. & SM & Original & Tapered & SM & Original & Tapered & SM & Original & Tapered & SM & Original & Tapered\\ \hline
      18 & -11.932 & -11.932 & -11.932 & -10.853 & -10.853 & -10.853 & -9.458 & -9.455 & -9.455 & -10.740 & -10.740 & -10.740\\
      20 & -23.632 &  -23.146 & -22.957 & -21.865 & -21.630 & -21.484 & -19.228 & -19.080 & -18.940 & -22.271 & -22.130 & -21.972 \\
      22 & -34.498 & -33.931 & -33.926 & -33.192 & -32.737 & -32.722 & -29.896 & -29.533 & -29.480 & -34.937 & -34.555 & -34.465 \\
      24 & -41.225 & -41.022 & -41.021 & -44.132 & -43.906 & -43.905 & -40.974 & -40.772 & -40.772 & -48.158 & -47.918 & -47.916 \\
      26 & -40.869 & -40.869 & -40.869 & -40.102 & -40.102 & -40.102 & -39.789 & -39.788 & -39.788 & -53.779 & -53.779 & -53.779 \\ \hline
    \end{tabular}    
    \end{adjustbox}

    \label{tab:summary}
\end{table*}

The quantum circuit design for the ground state of the $^{20}$O isotope is more complicated compared to the $^{18}$O circuit. It involves fourteen parameters, and the ground state energies obtained for different interactions possess 1.35\% to 2.86\% errors compared to the corresponding shell model results. This is due to considering only fifteen SD states in the construction of ansatz instead of 81 SD states to make the circuit simple enough so that it would be possible to run on quantum hardware. 
Our simulator result for $^{20}$O ground state energy using USDB interaction possesses an error ratio of 2.86\% which is slightly better than the result reported in \cite{QC_n3} for a ten-parameter ansatz. 
The circuit design for $^{24}$O isotope is similar to $^{20}$O having the same number of parameters. However, the error ratios are less than 0.52\% for all four interactions, which is much less compared to the $^{20}$O results.

The circuit design of $^{22}$O isotope is more difficult than the rest of the oxygen isotopes considered in this work. As it is a mid-shell nucleus, the $M$-scheme dimension is the highest for the ground state compared to the other four oxygen isotopes. We considered twenty SD states with 19 trainable parameters for the $^{22}$O ground state. While the error ratios for this isotope are between 1.35 \% to 1.66 \% for different interactions, the error ratio is only 0.37 \% for a 35-parameter circuit reported in Ref. \cite{QC_n3}.

From the above discussion, we can see that due to the simple structure of our quantum circuits, the simulated results are slightly less bound compared to the exact results. However, the trend of ground state energies is nicely reproduced by those circuits as can be seen in \autoref{simulator_summary}. While the phenomenological USDB interaction along microscopic interactions DJ16 and JISP16 establish $^{24}$O as the most bound nucleus, the N3LO interaction fails to do so.
 The lack of $3N$ force in the N3LO interaction is the main reason for which there is a significant mismatch between the experimental and calculated results. One way to improve the results of N3LO interaction for $^{24-26}$O would be to include a $3N$ contribution from the next to next to leading order (N2LO) of the chiral perturbation series. The other two microscopic interactions, namely JISP16 and DJ16, have been tuned to fit some selected properties of a few low-mass nuclei up to A = 16 without explicitly using the $3N$ force.

\subsection{Results from hardware}

In order to test the viability of our methods, we executed the circuits on real quantum hardware for all 5 isotopes at the variational minimum for the DJ16 interaction. The optimal parameters were first obtained through ideal classical simulation. We used IonQ's Aria, a trapped-ion quantum computer accessed through Microsoft's Azure Quantum cloud platform. A trapped-ion device was chosen because the 5-qubit tapered circuits necessitate all-to-all qubit connectivity. At the time of execution, average gate fidelities were reported on the Azure Quantum provider information page as 99.95\% for single-qubit gates, and 99.6\% for the two-qubit gate \cite{azure-docs}. 

These gate error rates necessitate the use of error mitigation. However, to enable this, it was necessary to bypass the compiler by submitting circuits to the cloud platform expressed in IonQ's native gate set. The set has three elements, two single-qubit gates \cite{native-gates},
\begin{eqnarray}
    GPI(\phi) &=& \begin{pmatrix}
     0 & e^{-i \phi} \\ e^{i \phi} & 0
    \end{pmatrix}, \\
    GPI2(\phi) &=& \frac{1}{\sqrt{2}} \begin{pmatrix}
        1 & -i e^{-i\phi} \\
        -i e^{i\phi} & 1
    \end{pmatrix},
\end{eqnarray}
\noindent and the two-qubit gate M{\o}lmer-S{\o}renson gate,
\begin{equation}
      MS = \frac{1}{\sqrt{2}} \begin{pmatrix}
        1 & 0 & 0 & -i \\
        0 & 1 & -i & 0 \\
        0 & -i & 1 & 0 \\
        -i & 0 & 0 & 1
    \end{pmatrix}.
\end{equation}

We wrote a tool to perform transpilation and optimization of PennyLane circuits into this gate set, which we make available open-source to the community \cite{ionizer}. Prior to transpilation, the circuits were also passed through the Qiskit transpiler with optimization level 3, in order to minimize the 2-qubit gate count. The resources for the resultant circuits are presented in \autoref{tab:resest-tapered}. For comparison, we also provide  estimates for the circuits transpiled to a connectivity-restricted processor (which requires the addition of many two-qubit SWAP gates for qubit routing) to further motivate the use of the trapped-ion device.

\begin{table}
    \centering
    \caption{Resource counts (1- and 2-qubit gate counts and circuit depth) for tapered ansatze. All circuits act only on 5 qubits. The original circuit is expressed in terms of $RZ$, $RY$, and $CNOT$ gates and was optimized using Qiskit's transpiler with optimization level 3. The trapped-ion version is obtained by transpiling and optimized the original circuits into the $GPI$, $GPI2$, $MS$ gate set using our custom transpiler \cite{ionizer}. The final column is the original circuit but placed and routed on the hardware graph of IBM's 7-qubit superconducting (SC) Nairobi device, whose topology is shaped like an ``H". Placement and routing is done with Qiskit using SABRE \cite{sabre}. SC results are provided only to show the effect of restricted topology.}
    \begin{tabular}{|c||c|c|c||c|c|c||c|c|c|}
      \hline
      & & Orig. & & & Ion & & & SC & \\ \hline
      Iso. & 1Q & 2Q & $d$ & 1Q & 2Q & $d$ & 1Q & 2Q & $d$\\ \hline
      18 &  40  & 8 & 24 & 46 & 8 & 26 & 40 & 8 &  24\\
      20 &  55 & 26 & 45 & 87 & 26 & 67 & 53 & 40 & 58 \\
      22 & 59  & 35  & 55 & 97 & 35 & 83 & 64 & 57 & 90 \\
      24 & 67 & 36 & 58  &  94 & 36 & 85 & 61 & 55 & 72 \\
      26 &  39 & 8 & 24 &  46 & 8 &  26 &  39 & 8 & 24 \\ \hline
 \end{tabular} 
    \label{tab:resest-tapered}
\end{table}

The initial hardware results are shown in the ``HW (raw)" column of \autoref{tab:hw-dripline-results}. They are the result of executing 8 circuits (one per qubit-wise commuting set of Paulis), with 1000 shots each, to compute the expectation values of the tapered Hamiltonian discussed in \autoref{subsec:tapering}. The results for the smallest problem instances ($^{18}$O and $^{26}$O) are within a few percent of the exact values, while there is significant deviation for $^{20}$O, $^{22}$O, and $^{24}$O, which require $\sim$4x as many two-qubit gates.

To improve the results we applied error mitigation. As the ions on hardware are not individually addressable through the software interface, measurement error mitigation was not performed since we could not tie results of calibration circuits to specific qubits with certainty. Instead, only gate error mitigation with zero-noise extrapolation (ZNE) \cite{zne} was performed for the three middle isotopes. Two-qubit gate folding was used. There are two ways to fold on the hardware platform: $MS$ folding after transpilation to native gates, or $CNOT$ folding prior to transpilation. For the former, $MS$ is not its own inverse ($MS^\dagger = M^3$). A single fold thus consists of 4 $MS$ gates, which adds substantially more noise, as well as financial cost to the simulation. 

\begin{figure*}
    \centering
    \includegraphics[width=0.8\textwidth]{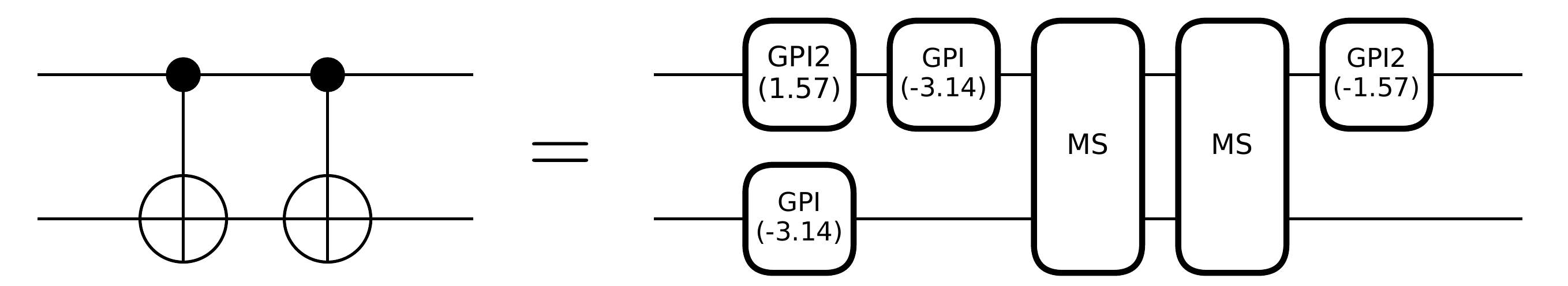}
    \caption{A $CNOT$ pair, to be inserted in error mitigation protocols, transpiled into trapped ion gates.}
    \label{fig:ms_fold}
\end{figure*}

$CNOT$, on the other hand, is its own inverse, so a fold consists simply of adding two of them. Typical extrapolation is performed with respect to noise scale factors of the form $2\lambda + 1$, where $\lambda$ is the number of folds, which scales directly with the number of additional $CNOT$ gates. After transpilation to trapped-ion gates, we obtain two MS gates and some single-qubit gates as shown in \autoref{fig:ms_fold}. In principle, the noise scale factor could be adjusted to take into account the additional single-qubit gates. However, the single-qubit error rates are a factor of 10 less and in many cases, they end up being absorbed into neighbouring gates during the optimization process, so we do not consider this. 

\begin{figure}
    \centering
    \includegraphics[width=0.5\textwidth]{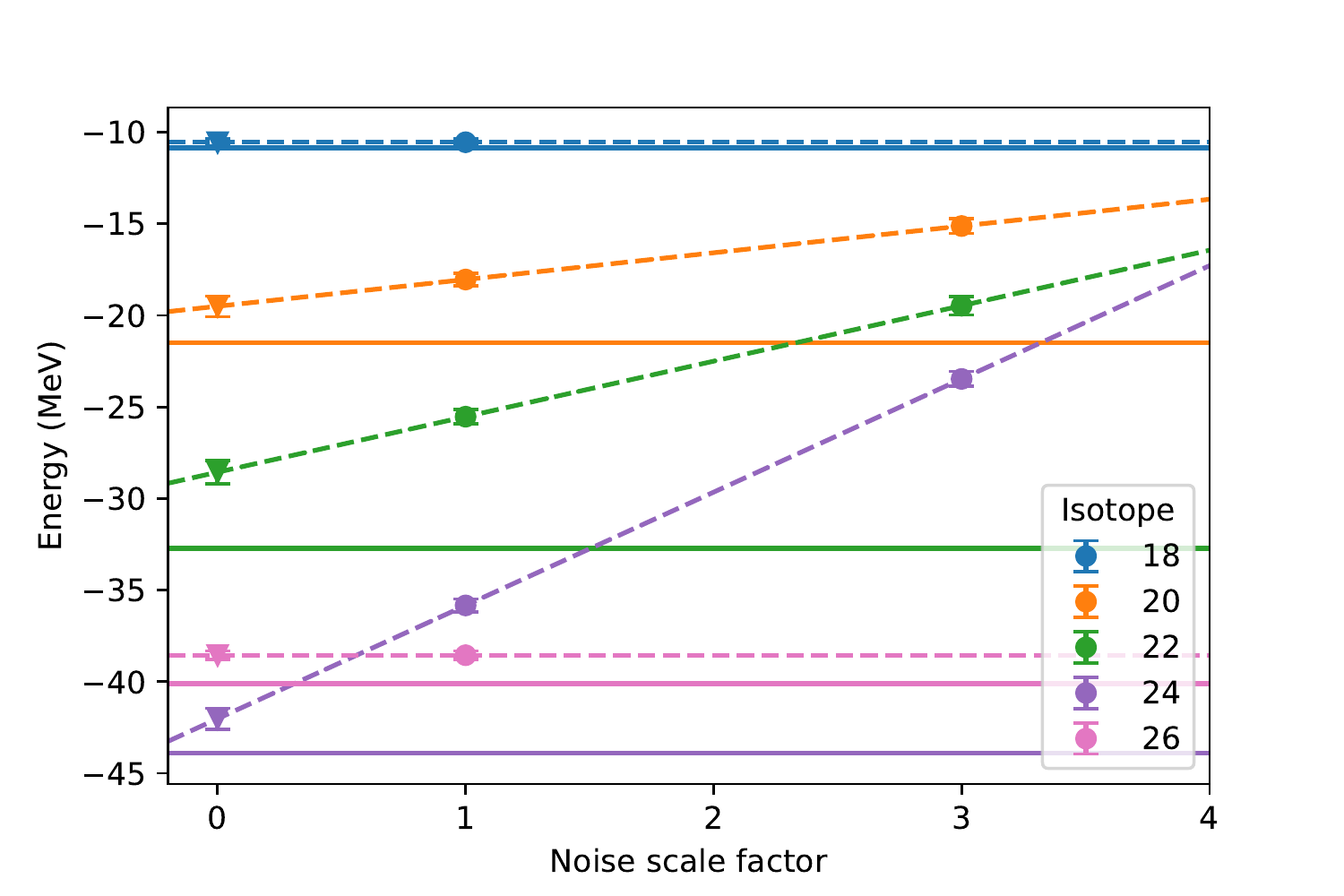}
    \caption{Results from hardware execution on IonQ Aria. Solid lines correspond to exact values. Raw results are shown at noise scale factor 1. The raw results alone do not show the drip line; to improve the results, zero-noise extrapolation with a single $CNOT$ fold and linear fit was performed for isotopes 20, 22, and 24. These results are shown at noise scale factor 3, and the dotted lines show the extrapolation back to the zero-noise limit. Results from isotopes 18 and 26 are repeated at 0 to improve visibility of the key result: after error mitigation, the drip line is visible. The error bars correspond to standard deviations estimated from 100 Monte Carlo simulations, bootstrapped from the raw results (see Appendix C).}
    \label{fig:aria-results}
\end{figure}

\begin{figure}
    \centering
    \includegraphics[width=0.54\textwidth]{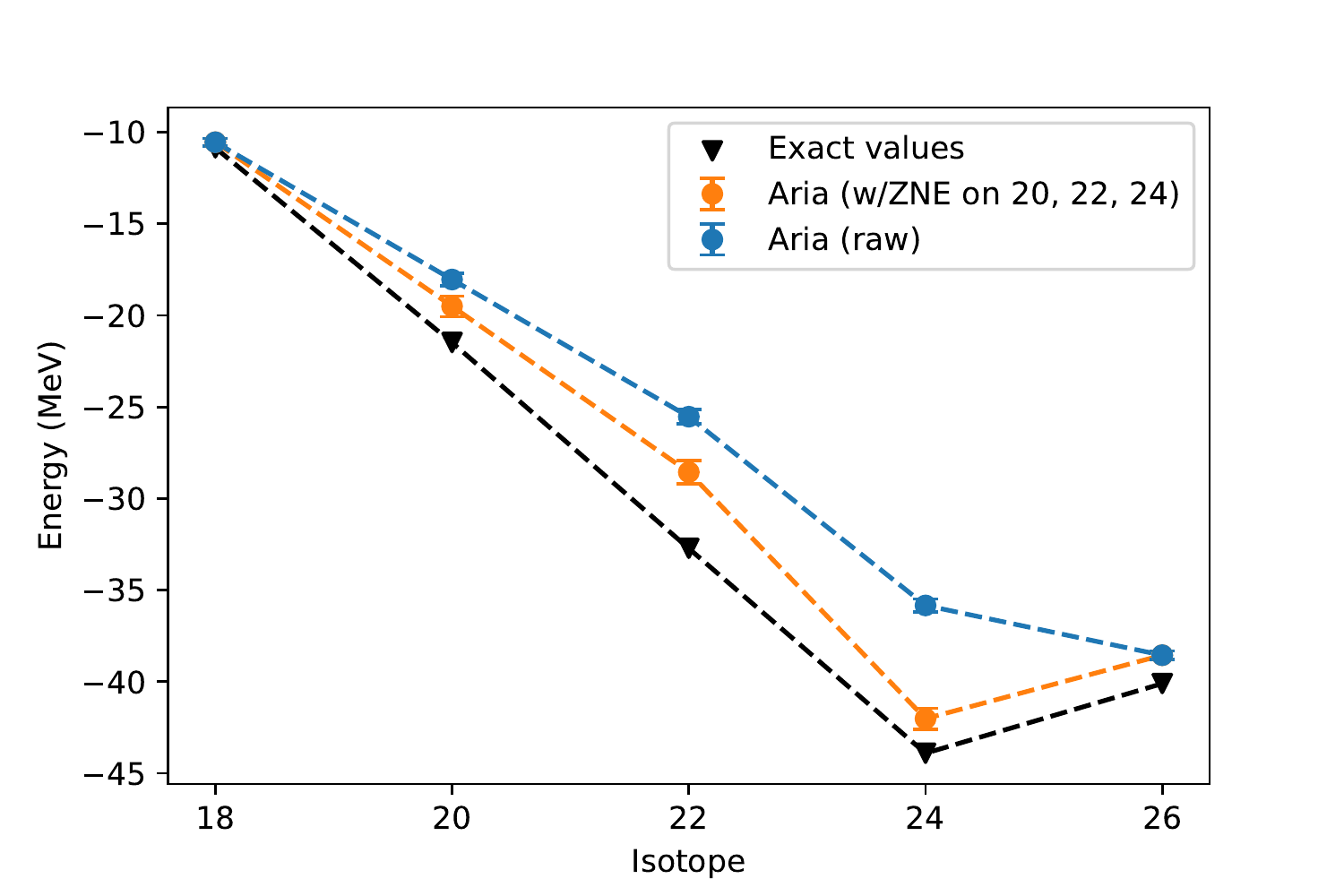}
    \caption{Plot of numerical results from \autoref{tab:hw-dripline-results}. The error bars correspond to standard deviations estimated from 100 Monte Carlo simulations, bootstrapped from the raw hardware results (see Appendix C). The drip line is clearly visible.}
    \label{fig:hw-dripline-plot}
\end{figure}

Due to resource constraints, only a single $CNOT$ fold was performed, and the resulting circuits were transpiled and optimized. The ZNE circuits were also executed with 1000 shots, and the extrapolation procedure was a simple linear fit of the final expectation value, which is shown in \autoref{fig:aria-results}. Since the fit includes only two points at consistent scale factors, we can perform it on the final Hamiltonian expectation values directly instead of on each Pauli term individually. The error rates of the hardware were low enough that even minimal ZNE proved sufficient to see the drip line; our final results are shown in \autoref{fig:hw-dripline-plot}. However, we note that the percentage error of two of the isotopes ($^{20}$O and $^{22}$O) is relatively higher than the rest. 

The error bars in the plots represent the standard deviation of Monte Carlo simulations bootstrapped from the probability distributions obtained from hardware execution. The bootstrapping procedure is described in detail in Appendix C. Future work should test the reliability and consistency of the results by performing more experiments on hardware, with more shots, more gate folds,  different extrapolation functions, or alternative mitigation techniques.

The results do, however, highlight the significant progress that has been made in quantum hardware over the past few years. As a point of comparison, ZNE with Richardson extrapolation was performed on a trapped-ion machine for the first time in \cite{shehab2019}  to solve for the deuteron binding energy using circuits with up to 4 qubits. Our baseline 5-qubit circuits for $^{22}$O and $^{24}$O use as many two-qubit gates as the largest circuits that were run in \cite{shehab2019}, which were the terminal points in the extrapolation. Even with minimal post-processing, we find percentage errors to be comparable to the state-of-the-art at the time (2-4\%). Additional post-processing, such as measurement error mitigation, would improve these results further, perhaps even bypassing the need for gate error mitigation.

\begin{table}
 \centering
 \caption{Numerical results from IonQ Aria machine at variational minimum for DJ16 interaction. All circuits were executed using 1000 shots. Zero-noise extrapolation with a single $CNOT$ fold and linear extrapolation was performed for isotopes 20, 22, and 24. The percent error is computed from the bold-faced value for all cases.}
 \begin{tabular}{|c|c|c|c|c|}
 \hline
    Iso. & Exact & HW (raw) & HW (+ZNE) & Percent error \\ \hline
    18 & -10.853 & {\bf -10.546} & $\cdot$  & 2.83 \\
    20 & -21.484 & -18.041 & {\bf -19.516} & 9.16 \\
    22 & -32.722 & -25.537 & {\bf -28.611} & 12.56 \\
    24 & -43.905 & -35.830 & {\bf -42.001} & 4.34 \\
    26 & -40.102 & {\bf -38.547} &  $\cdot$ & 3.88 \\ \hline
 \end{tabular}
 \label{tab:hw-dripline-results}
\end{table}

\section{Conclusions}

We investigated the usefulness of quantum computation in the shell model description of even oxygen isotopes from $N$ = 10 to 18 in the $sd$ valence space above a $^{16}$O core. Initially, we represented this problem as a 12-qubit problem as there are 12 single neutron states in the $sd$ space. We constructed UCCD type ansatze for solving the problem with the VQE, but the quantum resources required for running those circuits on quantum hardware are far away from the currently available resources. So, we use qubit tapering technique that reduces our 12-qubit problem to an equivalent 5-qubit problem. The quantum resources were reduced significantly so it became possible to run on quantum hardware. The classical simulator results for the tapered ansatze are in good agreement with the exact shell model diagonalization results. The simulated ground state energies of the oxygen isotopes for USDB, DJ16, and JISP16 follow the experimental trend and thus predict the neutron drip line for the oxygen chain correctly at $^{24}$O. However, the N3LO interaction fails to follow the same.

Finally, we run the 5-qubit quantum circuits on trapped-ion quantum hardware, IonQ-Aria using the optimal parameters from classical simulator. Due to limited quantum resources, we ran our quantum circuits only for the DJ16 interaction. The raw (without error-mitigation) results for low-depth circuits of $^{18,26}$O are close to the exact results within a few percent errors. However, the mid-shell oxygen isotopes, 20, 22, and 24 needed error-mitigation techniques to correctly reproduce the experimental trend of binding energies. The unmitigated results of the former two isotopes along with the zero-noise extrapolation (ZNE) results of the latter three isotopes show $^{24}$O as the neutron drip line nucleus. 

In the future, this work can be extended to include higher mass isotopic chains across the $sd$ model space like neon and magnesium. Due to the open shell nature of these two isotopic chains, the $M$-scheme Hamiltonian would contain $pp$, $nn$, and $pn$ parts, unlike oxygen where only $nn$ part of the Hamiltonian was sufficient. It will increase the number of Pauli strings significantly after the JW transformation.  Apart from the Hamiltonian, scaling up the quantum circuit design for these isotopic chains will be challenging within the UCC formalism as many double, single-controlled double, and double-controlled double excitation gates would be needed to construct suitable ansatze. The concurrent increase in the number of variational parameters may also lead to barren plateaus and challenges with trainability. To avoid such difficulties it would be important to explore these problems using automated ansatz construction techniques like ADAPT-VQE \cite{adapt-vqe, adapt-barren-plateaus} (as was done in recent work \cite{QC_n6}), or a Hamiltonian variational ansatz \cite{hva}.

\section*{ACKNOWLEDGMENTS}
We acknowledge financial support from SERB (India), Grant No. CRG/2022/005167, MHRD (India), NSERC (Canada) Grant No. RGPIN-2022-04609, NSERC CREATE program, and the Canada Research Chairs, Grant No. CRC-2021-00206 program. We acknowledge support from
Microsoft’s Azure Quantum for providing credits and access to the IonQ Aria quantum hardware used in this paper.

\section*{Appendix A}
\label{sec:appendix}
The two-body matrix elements from $J$-scheme to $M$-scheme is transformed using the following formula \cite{suhonen}:
\begin{eqnarray*}
    \bar{v}_{\alpha \beta \gamma \delta} = \sum_{J, M, T, T_z} [N_{ab}(JT) N_{cd}(JT)]^{-1} \left( j_a m_\alpha j_b m_\beta|JM\right) \hspace{10 cm}\\
    \times  \left( \frac{1}{2} m_{t \alpha} \frac{1}{2} m_{t \beta}|T M_T\right) \left( j_c m_\gamma j_d m_\delta|JM\right) \hspace{10 cm} \\
    \times \left(\frac{1}{2} m_{t \gamma} \frac{1}{2} m_{t \delta}|T M_T\right) \langle a b; J T | V | c d; J T \rangle. \hspace{9.9 cm}
\end{eqnarray*}

Here, the Greek letters $\alpha$, $\beta$, $\gamma$, $\delta$ represent single nucleon states as $|\alpha \rangle = |n, l, j, m_\alpha, t_z \rangle$. The four quantities inside bracket are the $3j$-symbols of angular momenta and isospins, while $\langle a b; J T | V | c d; J T \rangle$ are the $J$-scheme two-body matrix elements. The normalization constants $N_{ab}(JT)$ and $N_{cd} (JT)$ are defined as:

\begin{eqnarray*}
    N_{ab} (JT) = \frac{\sqrt{1-\delta_{\alpha \beta} (-1)^{J+T}}}{1+\delta_{\alpha \beta}},\\
    N_{cd} (JT) = \frac{\sqrt{1-\delta_{\gamma \delta} (-1)^{J+T}}}{1+\delta_{\gamma \delta}}.
\end{eqnarray*}

\section*{Appendix B}

\begin{figure}[h]
    \centering
    \includegraphics[width=7.5cm, trim=2cm 11cm 1cm 5cm, clip]{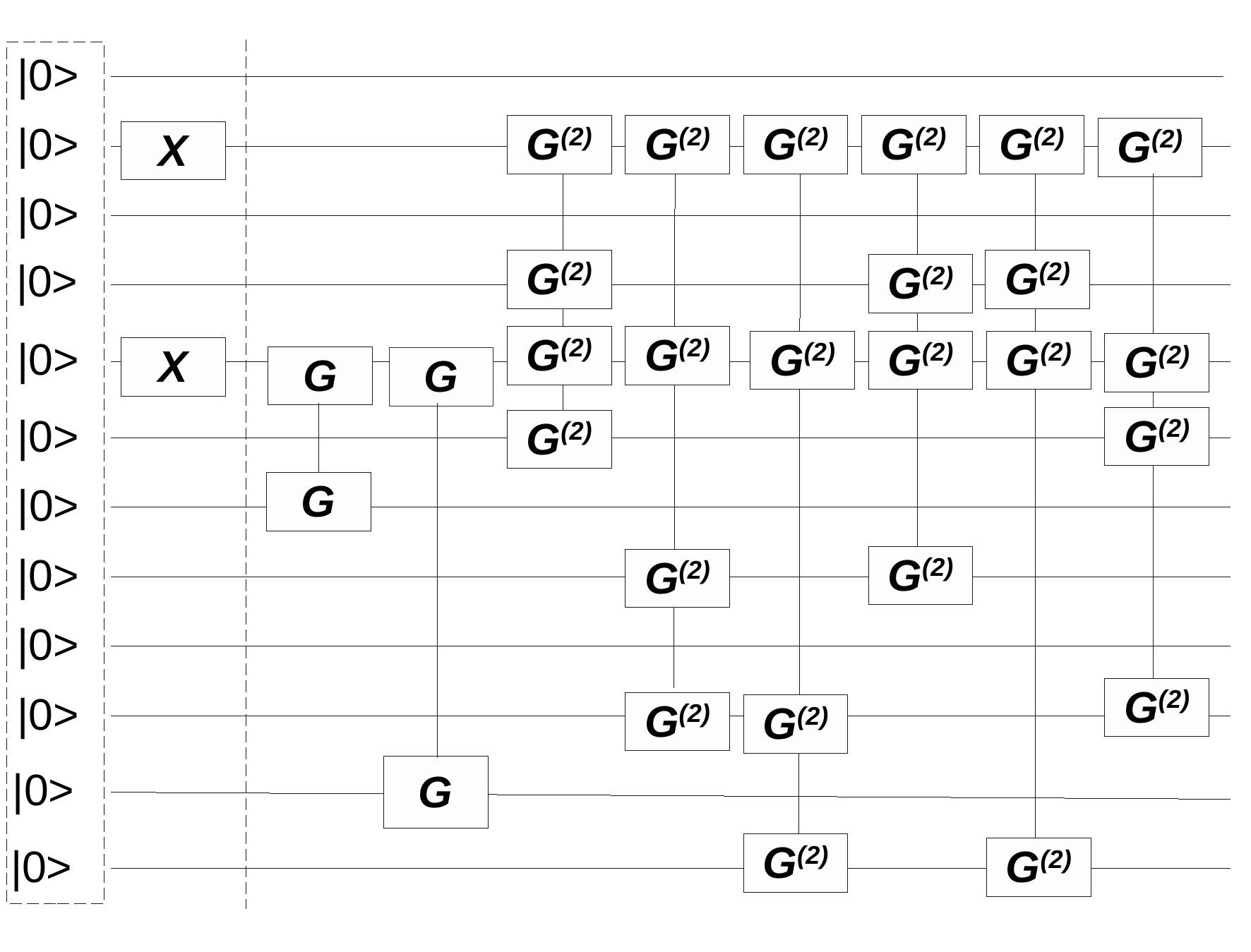}
    \caption{The first excited state ansatz of $^{18}$O considering $|1, 4\rangle$ as the reference states. The $G^{(2)}$ and $G$ represent the double and single excitation gates that are applied to get the desired number of SD states in the ansatze. Each single and double excitation gate adds one more parameter to the quantum circuit.}
    \label{18O_ansatz}
\end{figure}

The first excited state of $^{18}$O can be constructed using unitary coupled-cluster with singles and doubles (UCCSD) type ansatz using single, $G$, and double excitation gates $G^{(2)}$ acting on a reference state, as discussed in subsection II B. The construction of the circuit is shown in \autoref{18O_ansatz}.

 Here, the number of SDs considered for the construction of the ansatz is the same as the $M$-scheme dimension of that state. As shown in \autoref{18O_ansatz}, the ansatz is constructed by applying two single excitations and six double excitation gates on the reference state, $| 1, 4 \rangle$, to have an eight-parameter quantum circuit.

 \begin{figure}
    \centering
    \includegraphics[scale=0.5]{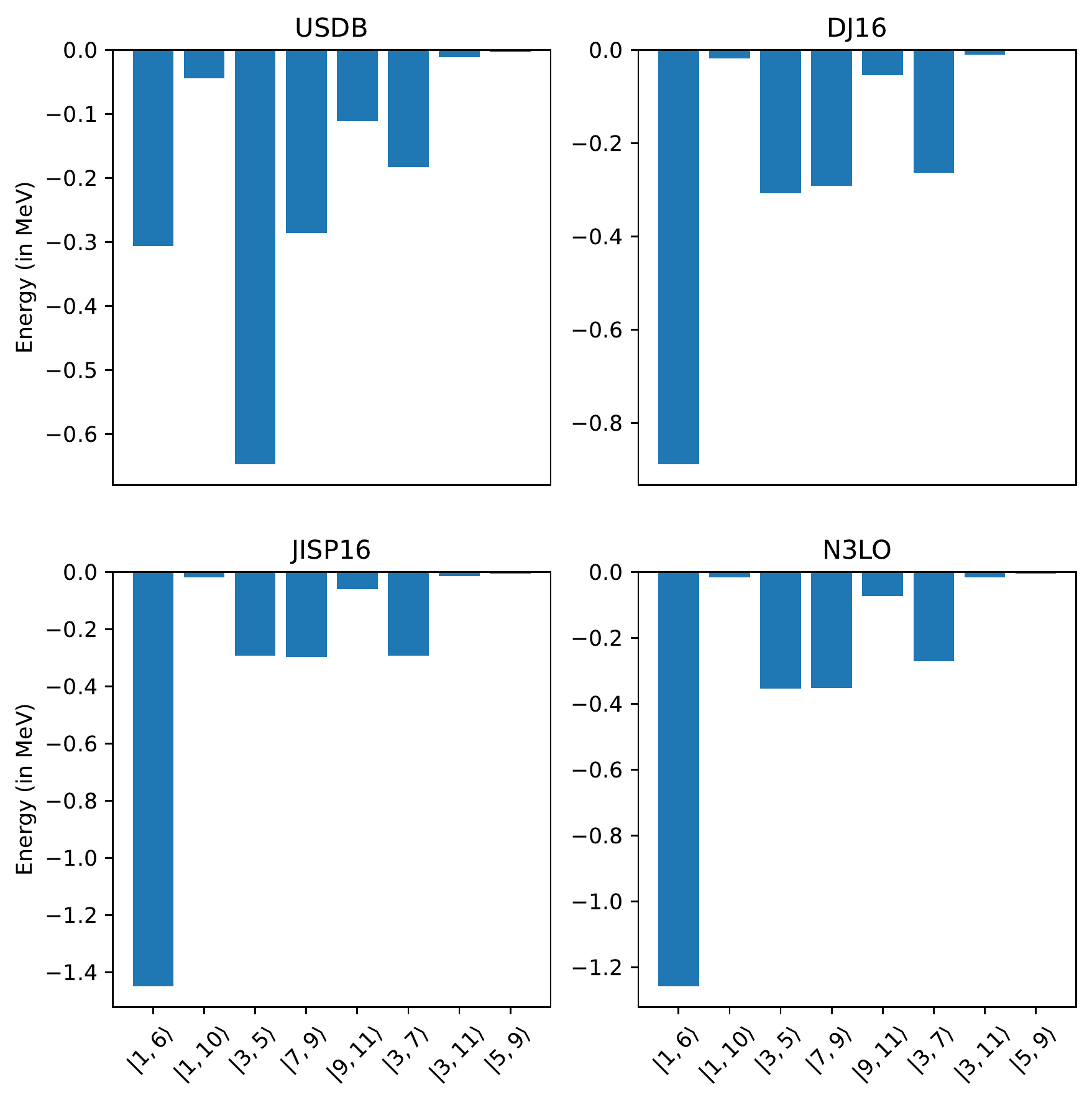}
   \caption{Energy contributions of different Slater determinant states to the binding energy of first excited state of $^{18}$O is shown. For all four cases, $|1, 4 \rangle$ is the reference state, and contributions from the reference state are different for different interactions.}    \label{fig:1st_excited}
\end{figure}

 As the number of SDs in the ansatz is the same as the $M$-scheme dimension of that state, the ideal simulator results match the exact results up to good approximation. The contributions from the different SDs apart from the reference state are shown in \autoref{fig:1st_excited}. From the figure, we can see that the contributions from $|3, 11\rangle$ and $|5, 9\rangle$ states are quite small, whereas the contributions from $|1, 6\rangle$ are the highest corresponding to the three microscopic interactions we considered in this work. This 12-qubit problem can be reduced to a 5-qubit problem using the same qubit tapering technique as used for the ground state. However, due to the limited quantum resources, we did not intend to run these circuits on quantum hardware. Similarly, it is possible to construct an ansatz for some of the low-energy states of other oxygen isotopes.

\section*{Appendix C}
The statistical uncertainty of the ground state energies obtained by executing the optimized VQE circuits on hardware can ideally be estimated by repeating the hardware experiments multiple times and computing the standard deviation. However, this is infeasible in practice due to resource constraints. In order to obtain a rough estimate of the standard deviation, we used Monte Carlo simulation bootstrapped using results of the experiment for each isotope.

A single experiment for a given isotope comprised of executing 8 circuits (one per qubit-wise commuting set of Paulis), with 1000 shots each. The data for each experiment (circuit) is returned from the hardware provider as a histogram where the bins and frequencies correspond to individual bitstrings and their observed frequencies respectively (for the 5-qubit VQE circuits used in this work, there was a maximum of $2^{5} = 32$ bins per histogram).

To estimate the standard deviation of the ground state energy value for a single isotope, we first used bootstrapped Monte Carlo simulations to independently sample from the distributions represented by each of the 8 histograms $n_{\text{shots}} (1000)$ times and post-processed the results to obtain a ground state energy value. We then repeated this process multiple $(100)$ times to obtain several ground state energy values with statistical fluctuations due to the finite number of Monte Carlo samples. The standard deviation in these values is used as an estimate of the true standard deviation of the ground state energy value obtained from the hardware.

We performed this procedure for different isotopes independently to obtain the standard deviations reported in \autoref{tab:mc-sim-std-dev}, which are those shown as error bars in \autoref{fig:aria-results} and \autoref{fig:hw-dripline-plot}. The same was done for the three isotopes where gate folding was applied. For the estimate of the zero-noise extrapolated values, we used a combination of the standard deviation estimates from the experiments with noise scale factors of 1 and 3 (computed as the square root of the combined variance obtained by taking a linear combination of the individual variances with squared coefficients).

We note that the key limitation of the above method is that it makes the simplifying assumptions that the device distribution is stationary, i.e., the device operational behaviour does not change over time (which is unlikely to hold for near-term noisy devices), and that the number of measurement shots in the original experiments are sufficient to capture the bitstring distributions.

\begin{table}
 \centering
 \caption{Monte Carlo estimates of the standard deviation of the ground state energy values (in MeV) for different isotopes. Bootstrapped Monte Carlo simulation using single experiment data was used to estimate the std. dev. for noise scale factors of 1 and 3, while a linear combination of the two was used to estimate the std. dev. in the ZNE extrapolated ground state energy values for three of the isotopes.}
 \begin{tabular}{|c|c|c|c|}
 \hline
    Iso. & ZNE & Noise scale & Noise scale \\
     & extrapolated & factor = 1 & factor = 3 \\ \hline
    18 & $\cdot$ & 0.217 & $\cdot$ \\
    20 & 0.592 & 0.35 & 0.4 \\
    22 & 0.632 & 0.384 & 0.502\\
    24 & 0.497 & 0.357 & 0.403\\
    26 & $\cdot$ & 0.235 & $\cdot$ \\ \hline
 \end{tabular}
 \label{tab:mc-sim-std-dev}
\end{table}


\newpage

\bibliographystyle{utphys}
  \bibliography{references}

\begin{thebibliography}{44}
\expandafter\ifx\csname natexlab\endcsname\relax\def\natexlab#1{#1}\fi
\expandafter\ifx\csname bibnamefont\endcsname\relax
  \def\bibnamefont#1{#1}\fi
\expandafter\ifx\csname bibfnamefont\endcsname\relax
  \def\bibfnamefont#1{#1}\fi
\expandafter\ifx\csname citenamefont\endcsname\relax
  \def\citenamefont#1{#1}\fi
\expandafter\ifx\csname url\endcsname\relax
  \def\url#1{\texttt{#1}}\fi
\expandafter\ifx\csname urlprefix\endcsname\relax\def\urlprefix{URL }\fi
\providecommand{\bibinfo}[2]{#2}
\providecommand{\eprint}[2][]{\url{#2}}





\bibitem{SM1} B. A. Brown, ``The nuclear shell model towards the drip lines'',
\href{https://doi.org/10.1016/S0146-6410(01)00159-4} 
 { Prog. Part. Nucl. Phys. {\bf 47}, 517 (2001).}
 
\bibitem{SM2} E. Caurier, G. Martínez-Pinedo, F. Nowacki, A. Poves, and A. P. Zuker, ``The shell model as a unified view of nuclear structure'', \href{https://doi.org/10.1103/RevModPhys.77.427}{Rev. Mod. Phys. {\bf 77} 427 (2005).}

\bibitem{SM3} T. Otsuka, A. Gade, O. Sorlin, T. Suzuki, and Y. Utsuno, ``Evolution of shell structure in exotic nuclei'', \href{https://journals.aps.org/rmp/abstract/10.1103/RevModPhys.92.015002}{Rev. Mod. Phys. {\bf 92}, 015002 (2020).}

\bibitem{nisq} J. Preskill, ``Quantum Computing in the NISQ era and beyond'', \href{https://doi.org/10.22331/q-2018-08-06-79}{Quantum {\bf 2}, 79 (2018).}

\bibitem{vqe} A. Peruzzo, J. McClean, P. Shadbolt et al. ``A variational eigenvalue solver on a photonic quantum processor'', \href{https://doi.org/10.1038/ncomms5213}{Nat. Commun. 5, 4213 (2014).}

\bibitem{QC_n1} E. F. Dumitrescu, A. J. McCaskey, G. Hagen, G. R.
Jansen, T. D. Morris, T. Papenbrock, R. C. Pooser, D. J.
Dean, and P. Lougovski, ``Cloud Quantum Computing of an Atomic Nucleus'', \href{https://doi.org/10.1103/PhysRevLett.120.210501}{Phys. Rev. Lett. 120, 210501 (2018).}

\bibitem{QC_n2} H.-H. Lu, N. Klco, J. M. Lukens, T. D. Morris, A. Bansal,
A. Ekström, G. Hagen, T. Papenbrock, A. M. Weiner,
M. J. Savage, and P. Lougovski, ''Simulations of subatomic many-body physics on a quantum frequency processor'', \href{https://doi.org/10.1103/PhysRevA.100.012320}{Phys. Rev. A {\bf 100}, 012320 (2019).}

\bibitem{graycode} O. Di Matteo, A. McCoy, P. Gysbers, T. Miyagi, R. M. Woloshyn, and P. Navrátil, ``Improving Hamiltonian encodings with the Gray code'', \href{https://doi.org/10.1103/PhysRevA.103.042405}{Phys. Rev. A 103, 042405 (2021).}

\bibitem{pooja1} P. Siwach and P. Arumugam, ``Quantum simulation of nuclear Hamiltonian with a generalized transformation for Gray code encoding'', \href{https://doi.org/10.1103/PhysRevC.104.034301}{Phys. Rev. C {\bf 104}, 034301 (2021)}

\bibitem{pooja2} P. Siwach and P. Arumugam, `` Quantum computation of nuclear observables involving linear combinations of unitary operators'', \href{https://doi.org/10.1103/PhysRevC.105.064318}{Phys. Rev. C {\bf 105}, 064318 (2022).}

\bibitem{pooja3} P. Siwach and D. Lacroix, ``Filtering states with total spin on a quantum computer'', \href{https://doi.org/10.1103/PhysRevA.104.062435}{Phys. Rev. A 104, 062435 (2021).}

\bibitem{QC_n3} I. Stetcu, A. Baroni, and J. Carlson, ``Variational approaches to constructing the many-body nuclear ground state for quantum computing'', \href{https://doi.org/10.1103/PhysRevC.105.064308}{Phys. Rev. C {\bf 105}, 064308 (2022).}

\bibitem{QC_n4} A. M. Romero, J. Engel, H. L. Tang, and S. E.
Economou, ``Solving nuclear structure problems with the adaptive variational quantum algorithm'', \href{https://doi.org/10.1103/PhysRevC.105.064317}{ Phys. Rev. C {\bf 105}, 064317 (2022).}

\bibitem{QC_n5} O. Kiss, M. Grossi, P. Lougovski, F. Sanchez, S. Vallecorsa, and T. Papenbrock, ``Quantum computing of the $^6$Li nucleus via ordered unitary coupled clusters'', \href{https://doi.org/10.1103/PhysRevC.106.034325}{Phys. Rev. C {\bf 106}, 034325 (2022)}.

\bibitem{QC_n6} A. Pérez-Obiol, A. M. Romero, J. Menéndez, A. Rios, A. García-Sáez, B. Juliá-Díaz, ``Nuclear shell-model simulation in digital quantum computers'', \href{https://doi.org/10.48550/arXiv.2302.03641}{arXiv:2302.03641.}

\bibitem{QC_n7} I. Hobday, P. D. Stevenson, and J. Benstead, ``Quantum Computing Calculations for Nuclear Structure and Nuclear Data'', \href{https://doi.org/10.48550/arXiv.2205.05576}{arXiv:2205.05576.}

\bibitem{QC_n8} W. Du, J. P. Vary, X. Zhao, and W. Zuo, ``Quantum simulation of nuclear inelastic scattering'', \href{https://doi.org/10.1103/PhysRevA.104.012611}{Phys. Rev. A {\bf 104}, 012611 (2021).}

\bibitem{QC_n9} D. Lacroix, ``Symmetry-Assisted Preparation of Entangled Many-Body States on a Quantum Computer'', \href{https://doi.org/10.1103/PhysRevLett.125.230502}{Phys. Rev. Lett. 125, 230502 (2020).}

\bibitem{dripline1} H. Sakurai, S.M. Lukyanov, M. Notani, N. Aoi, $et$ $al$., ``Evidence for particle stability of $^{31}$F and particle instability of $^{25}$N and $^{28}$O'', \href{https://doi.org/10.1016/S0370-2693(99)00015-5}{Phys. Lett. B, {\bf 448} (1999), p. 180}

\bibitem{dripline2} N. Tsunoda, T. Otsuka, K. Takayanagi \textit{et al},``The impact of nuclear shape on the emergence of the neutron dripline'',
\href{https://doi.org/10.1038/s41586-020-2848-x}{Nature {\bf 587}, 66–71 (2020).}


\bibitem{dripline3}D. S. Ahn \textit{et al.}, Location of the neutron dripline at fluorine and neon, 
\href{https://doi.org/10.1103/PhysRevLett.123.212501}{ Phys. Rev. Lett. {\bf 123}, 212501 (2019).}

\bibitem{39Na} D. S. Ahn \textit{et al.}, 
``Discovery of $^{39}$Na',
\href{https://doi.org/10.1103/PhysRevLett.129.212502}{ Phys. Rev. Lett. {\bf 129}, 212502 (2022).}

\bibitem{antoine} E. Caurier and F. Nowacki, ``Present Status of Shell Model Techniques'', \href{https://www.actaphys.uj.edu.pl/R/30/3/705}{Acta Physica Polonica {\bf 30}, 705 (1999).}

\bibitem{nushellx} B. Brown and W. Rae, ``The Shell-Model Code NuShellX@MSU'',
\href{https://doi.org/10.1016/j.nds.2014.07.022}{Nuclear Data Sheets {\bf 120}, 115 (2014).}

\bibitem{kshell} 
N. Shimizu, T. Mizusaki, Y. Utsuno and Y. Tsunoda,
``Thick-restart block Lanczos method for large-scale shell-model calculations'', 
\href{Comput. Phys. Comm. 244, 372 (2019).}{Phys. Comm. {\bf 244}, 372 (2019).}

\bibitem{Bigstick} 
C. W. Johnson, W.E. Ormand, and P.G. Krastev,
``Factorization in large-scale many-body calculations'', 
\href{Comput. Phys. Comm. 184, 2761 (2013).}{Phys. Comm. {\bf 184}, 2761 (2013).}

 
\bibitem{usdb} B. A. Brown and W. A. Richter, ``New ``USD'' Hamiltonians for the $sd$ shell'', \href{http://dx.doi.org/10.1103/PhysRevC.74.034315}{Phys. Rev. C {\bf 74}, 034315 (2006).}
 
\bibitem{sd_int1} E. Dikmen, A. F. Lisetskiy, B. R. Barrett, P. Maris, A.M. Shirokov, J.P. Vary, ``$Ab$ $initio$ effective interactions for sd-shell valence nucleons'', \href{https://doi.org/10.1103/PhysRevC.91.064301}{Phys. Rev. C {\bf91}, 064301 (2015).} 

\bibitem{sd_int2} 
N. A. Smirnova, B. R. Barrett, Y. Kim, I. J. Shin, A. M. Shirokov, E. Dikmen, P. Maris, and J. P. Vary, ``Effective interactions in the $sd$ shell'', \href{https://doi.org/10.1103/PhysRevC.100.054329}{Phys. Rev. C {\bf 100}, 054329 (2019).}



\bibitem{jisp16} A. Shirokov, A. Mazur, J. Vary, and T. Weber, ``Realistic nuclear Hamiltonian: $Ab$ $exitu$ approach'', 
\href{https://doi.org/10.1016/j.physletb.2006.10.066}{Phys. Lett. B 644, 33 (2007).}

\bibitem{n3lo} D. R. Entem and R. Machleidt, ``Accurate charge-dependent nucleon-nucleon potential at fourth order of chiral perturbation theory'',
\href{https://doi.org/10.1103/PhysRevC.68.041001}{Phys. Rev. C 68, 041001(R) (2003).}

\bibitem{dj16} A. M. Shirokov, I. J. Shin, Y. Kim, M. Sosonkina, P. Maris, and J. P. Vary, ``N3LO NN interaction adjusted to light nuclei in $ab$ $exitu$ approach''
\href{https://doi.org/10.1016/j.physletb.2016.08.006}{Phys. Lett. B 761, 87 (2016).}

\bibitem{ncsm_2013} B. R. Barrett, P. Navrátil and J. P. Vary, ``Ab initio no
core shell model'', \href{https://doi.org/10.1016/j.ppnp.2012.10.003}{Prog. Part. Nucl. Phys. {\bf 69}, 131 (2013).}

\bibitem{priyanka1} P. Choudhary, P. C. Srivastava, and P. Navr\'atil, $Ab$ $initio$ no-core shell model study of $^{10-14}$B
isotopes with realistic $NN$ interactions, 
\href{https://doi.org/10.1103/PhysRevC.102.044309}
{Phys. Rev. C {\bf 102}, 044309 (2020).}

\bibitem{priyanka2} P. Choudhary, P. C. Srivastava, M. Gennari, and P. Navr\'atil, \textit{Ab initio} no-core shell-model description of $^{10-14}$C isotopes, 
\href{https://doi.org/10.1103/PhysRevC.107.014309}
{Phys. Rev. C  {\bf 117}, 014309 (2023).}

\bibitem{chandan} C. Sarma and P. C. Srivastava, \textit{Ab-initio} no-core shell model study of $^{18-24}$Ne isotopes, 
\href{https://doi.org/10.1088/1361-6471/acb962}
{J. Phys. G: Nucl. Part. Phys. {\bf 50}, 045105 (2023).}



\bibitem{ols_1994} K. Suzuki and R. Okamoto, ``Effective interaction theory
and unitary-model-operator approach to nuclear saturation problem'', \href{https://doi.org/10.1143/ptp/92.6.1045}{Prog. Theor. Phys. {\bf 92}, 1045 (1994).}

\bibitem{priyanka_EPJA}
P. Choudhary and P.C. Srivastava,
``Study of S, Cl and Ar isotopes with $N \geq Z$  using microscopic effective sd-shell interactions'', \href{https://doi.org/10.1140/epja/s10050-023-01013-8}{Eur. Phys. J. A {\bf 59}, 97 (2023).}

\bibitem{priyanka_NPA}
P. Choudhary and P.C. Srivastava,
``Nuclear structure properties of Si and P isotopes with the microscopic effective interactions'', \href{https://doi.org/10.1016/j.nuclphysa.2023.122629}{Nucl. Phys. A {\bf 1033}, 122629 (2023).}

\bibitem{subh_NPA}
S. Sahoo, P.C. Srivastava and T. Suzuki,
``Study of structure and radii for $^{20-31}$Na isotopes using microscopic interactions'', \href{https://doi.org/10.1016/j.nuclphysa.2023.122618}{Nucl. Phys. A {\bf 1032}, 122618 (2023).}

\bibitem{github} Code repository, \href{https://github.com/csarma24/Oxygen-chain-qc.git}{https://github.com/csarma24/Oxygen-chain-qc.git.}

\bibitem{vqa_review} M. Cerezo, A. Arrasmith, R. Babbush et al. ``Variational quantum algorithms.'' \href{https://doi.org/10.1038/s42254-021-00348-9}{Nat. Rev. Phys. {\bf 3}, 625–644 (2021).}

\bibitem{quantum_chem} S. McArdle, S. Endo, A. Aspuru-Guzik, S. C. Benjamin, and X. Yuan. ``Quantum computational chemistry'' \href{https://doi.org/10.1103/RevModPhys.92.015003}{Rev. Mod. Phys. {\bf 92}, 015003 (2020).}


\bibitem{universal_gates} J. M. Arrazola, O. Di Matteo, N. Quesada, S. Jahangiri, A. Delgado, and N. Killoran, `` Universal quantum circuits for quantum chemistry'', \href{https://doi.org/10.22331/Q-2022-06-20-742}{Quantum {\bf 6}, 742 (2022).}

\bibitem{qiskit} Qiskit contributions, ``Qiskit: An Open-source Framework for Quantum Computing'', \href{https://doi.org/10.5281/zenodo.2573505}{https://doi.org/10.5281/zenodo.2573505}. 


\bibitem{qtap1} S. Bravyi, J. M. Gambetta, A. Mezzacapo, K. Temme, ``Tapering off qubits to simulate fermionic Hamiltonians'', \href{https://doi.org/10.48550/arXiv.1701.08213}{arXiv:1701.08213.}

\bibitem{qtap2} K. Setia, R. Chen, J. E. Rice, A. Mezzacapo, M. Pistoia, J. Whitfield, ``Reducing Qubit Requirements for Quantum Simulations Using
Molecular Point Group Symmetries'', \href{https://doi.org/10.1021/acs.jctc.0c00113}{J. Chem. Theory Comput. 2020, 6091-6097.}

\bibitem{pennylane} V. Bergholm et al. ``PennyLane: Automatic differentiation of hybrid quantum-classical computations'', \href{https://doi.org/10.48550/arXiv.1811.04968}{arXiv:1811.04968}.

\bibitem{pennylane-qchem} J. M. Arrazola et al. ``Differentiable quantum computational chemistry with PennyLane'', \href{https://doi.org/10.48550/arXiv.2111.09967}{arXiv:2111.09967}.

\bibitem{elementary_gates} A. Barenco, C. H. Bennett, R. Cleve, D. P. DiVincenzo, N. Margolus, P. Shor, T. Sleator, J. A. Smolin, and H. Weinfurter. ``Elementary gates for quantum computation'', \href{https://doi.org/10.1103/PhysRevA.52.3457}{Phys. Rev. A {\bf 52}, 3457 (1995).}

\bibitem{templates} D. M. Miller, D. Maslov and G. W. Dueck, ``A transformation based algorithm for reversible logic synthesis,'' \href{https://doi.org/10.1145/775832.775915}{Proceedings 2003. Design Automation Conference (IEEE Cat. No.03CH37451), Anaheim, CA, USA, 2003, pp. 318-323}.

\bibitem{jax} J. Bradbury, R. Frostig, P. Hawkins, M. J. Johnson, C. Leary, D. Maclaurin, G. Necula, A. Paszke, J. Vander{P}las, S. Wanderman-{M}ilne, Q. Zhang, ``{JAX}: composable transformations of {P}ython+{N}um{P}y programs'', \href{https://github.com/google/jax}{http://github.com/google/jax}.

\bibitem{NNDC} Data extracted using the NNDC World Wide Web site from the ENSDF, 
\href{https://www.nndc.bnl.gov/ensdf/.}{ https://www.nndc.bnl.gov/ensdf/.}	

\bibitem{azure-docs} Microsoft Azure Quantum, ``IonQ Provider''. (Online) Accessed May 2023. \href{https://learn.microsoft.com/en-ca/azure/quantum/provider-ionq\#ionq-aria-quantum-computer}{https://learn.microsoft.com/en-ca/azure/quantum/provider-ionq\#ionq-aria-quantum-computer}

\bibitem{native-gates} IonQ Staff, ``Getting started with native gates''. (Online) \href{https://ionq.com/docs/getting-started-with-native-gates#introducing-the-native-gates}{https://ionq.com/docs/getting-started-with-native-gates}

\bibitem{ionizer} \href{https://www.github.com/QSAR-UBC/ionizer}{https://www.github.com/QSAR-UBC/ionizer} 

\bibitem{sabre} G. Li, Y. Ding, and Y. Xie, ``Tackling the Qubit Mapping Problem for NISQ-Era Quantum Devices'', \href{https://doi.org/10.48550/arXiv.1809.02573}{arXiv:1809.02573.}

\bibitem{zne} K. Temme, S. Bravyi, and J. M. Gambetta, ``Error Mitigation for Short-Depth Quantum Circuits'',
\href{https://doi.org/10.1103/PhysRevLett.119.180509}{Phys. Rev. Lett. {\bf 119}, 180509 (2017)}.

\bibitem{shehab2019} O. Shehab, K. Landsman, Y. Nam, D. Zhu, N. M. Linke, M. Keesan, R. C. Pooser, and C. Monroe, ``Toward convergence of effective-field-theory simulations on digital quantum computers'',
\href{https://doi.org/10.1103/PhysRevA.100.062319}{Phys. Rev. A {\bf 100}, 062319 (2019).}

\bibitem{adapt-vqe} H. R. Grimsley,  S.E. Economou, E. Barnes et al. ``An adaptive variational algorithm for exact molecular simulations on a quantum computer''. \href{https://doi.org/10.1038/s41467-019-10988-2}{Nat Commun {\bf 10}, 3007 (2019)}.

\bibitem{adapt-barren-plateaus} H. R. Grimsley,  G. S. Barron, E. Barnes et al. ``Adaptive, problem-tailored variational quantum eigensolver mitigates rough parameter landscapes and barren plateaus'',
\href{https://doi.org/10.1038/s41534-023-00681-0}{npj Quantum Inf {\bf 9}, 19 (2023).} 

\bibitem{hva} C.-Y. Park and N. Killoran. N. ``Hamiltonian variational ansatz without barren plateaus''. \href{https://doi.org/10.48550/arXiv.2302.08529}{arXiv:2302.08529.}

\bibitem{suhonen} J. Suhonen, ``From Nucleons to Nucleus: Concept of Microscopic Nuclear Theory, (Springer, Berlin 2007).


\textbf{}
\end{thebibliography}

\end{document}